\documentclass[12pt]{article}
\usepackage{multirow,axodraw}
\usepackage{amsmath,amsfonts}%,bbold}
\usepackage{amssymb}
\usepackage{hhline}
\usepackage{epsfig,cite}
\usepackage{stmaryrd}
\usepackage[usenames,dvips]{color}
\usepackage{fullpage}
\usepackage{verbatim}
   \allowdisplaybreaks
\makeatletter
\@addtoreset{equation}{section}
\makeatother

\newcommand{\be}{\begin{equation}}
\newcommand{\ee}{\end{equation}}
\newcommand{\bea}{\begin{eqnarray}}
\newcommand{\eea}{\end{eqnarray}}

\newcommand{\h}{\mathfrak h}

\begin{document}

\thispagestyle{empty}

\begin{center}
\hfill UAB-FT-744
\begin{center}

\vspace{.5cm}

{\Large\sc Higgs Bosons in Extra Dimensions$^\dagger$}

\end{center}

\vspace{1.cm}

\textbf{Mariano Quiros}\\

\vspace{.5cm}

 {\em Department of Physics, CERN Theory Division,
CH-1211, Geneva 23, Switzerland\\and\\
Instituci\'o Catalana de Recerca i Estudis  
Avan\c{c}ats (ICREA) \\ Institut de F\'isica d'Altes Energies, Universitat Aut{\`o}noma de Barcelona\\
08193 Bellaterra, Barcelona, Spain}

\end{center}

\vspace{0.8cm}

\centerline{\bf Abstract}
\vspace{2 mm}
\begin{quote}\small
In this paper, motivated by the recent discovery of a Higgs-like boson at the LHC with a mass $m_H\simeq 126$ GeV, we review different models where the hierarchy problem is solved by means of a warped extra dimension. In the Randall-Sundrum model electroweak observables provide very strong bounds on the mass of KK modes which motivates extensions to overcome this problem. Two extensions are briefly discussed. One particular extension is based on the deformation of the metric such that it strongly departs from the AdS$_5$ structure in the IR region while it goes asymptotically to AdS$_5$ in the UV brane. This model has the IR brane close to a naked metric singularity (which is outside the physical interval) characteristic of soft-walls constructions. The proximity of the singularity provides a strong wave-function renormalization for the Higgs field which suppresses the $T$ and $S$ parameters. The second class of considered extensions are based on the introduction of an extra gauge group in the bulk such that the custodial $SU(2)_R$ symmetry is gauged and protects the $T$ parameter. By further enlarging the bulk gauge symmetry one can find models where the Higgs is identified with the fifth component of gauge fields and for which the Higgs potential, along with the Higgs mass, can be dynamically determined by the Coleman-Weinberg mechanism.
 \end{quote}

\vline

\vfill
$^\dagger$\textit{Invited Review Paper submitted to Int.~J.~Mod.~Phys.~A special issue on "Extra Dimensions versus Collider Physics"}
 
 \newpage
\section{Introduction}
The discovery of the Higgs boson [the last missing ingredient of the Standard Model (SM)] at the LHC by the ATLAS~\cite{:2012gk} and CMS~\cite{:2012gu} collaborations, with a mass $m_H\simeq 126$ GeV, has now completed the Standard Model (SM) picture and made a big progress in clarifying the mechanism of electroweak symmetry breaking (EWSB). Although its nature, perturbative [as in the original Brout-Englert-Higgs (BEH) mechanism] versus non-perturbative [as in QCD], is not yet fully unveiled, everything seems to point towards a perturbative mechanism in the SM, although the accuracy of experimental data is not yet enough to rule out physics Beyond the Standard Model (BSM). Moreover from the theoretical point of view such electroweak vacuum would not be stable under quantum corrections (also known as hierarchy problem) which contribute to the Higgs mass with quadratically divergent radiative corrections, provided that we consider the SM as an effective theory. In particular the top quark contributes to the Higgs self-energy as
\begin{equation}
\Delta m_H^2=-\frac{3h_t^2}{8\pi^2}\Lambda^2
\label{top}
\end{equation}
where $\Lambda$ is the SM cutoff. In the absence of any tuning this implies an upper bound on the cutoff scale as
\begin{equation}
\Lambda\lesssim 3\, m_H
\end{equation}
and therefore beyond this cutoff there should exist new physics to stabilize it. Uncovering the nature of the electroweak symmetry breaking should amount to uncovering the kind of new physics (if any) which stabilizes the electroweak vacuum!

In fact there are two main avenues for solving the hierarchy problem depending on whether we consider the Higgs as an elementary scalar (perturbative EWSB) or a composite state (non-perturbative EWSB). The paradigm of perturbative EWSB which solves the hierarchy problem is supersymmetric theories, where there is an extra symmetry (supersymmetry) and extra matter (sfermions, gauginos and Higgsinos) with couplings dictated by supersymmetry such that there are extra contributions to the Higgs mass which cancel its quadratic sensitivity on the scale. For instance there is an extra scalar boson with the same quantum number of the top quark (the \textit{stop}) whose contribution cancels that of the quark top in Eq.~(\ref{top}). This is \textit{not} the solution we will cover in this review article.

The hierarchy problem is also solved provided that at some scale the Higgs dissolves, and the theory of its constituents is at work. This case is similar to QCD where the pions dissolve into quarks and gluons beyond $\Lambda_{QCD}$. In fact the compositeness scale acts as a cutoff of quadratic divergences. The typical example of compositeness is Technicolor. Modern theories of compositeness involve an extra dimension through the
Anti-de-Sitter/Conformal-Field-Theory (AdS/CFT) correspondence~\cite{Maldacena:1997re}. This provides an extra motivation to study models in a warped extra dimension,  the paradigm being the original five-dimensional (5D) Randall-Sundrum (RS) setup~\cite{Randall:1999ee} which, in particular, we will review in this article whose contents are as follows. 

In section 2 we will consider some general features of the 5D Standard Model propagating in the bulk of an extra dimension with an arbitrary warp factor. In section 3 we will provide general expressions for the oblique electroweak observables $S$, $T$, $W$ and $Y$ for a general warp factor and a general bulk Higgs profile. For completeness we will also provide general expressions for the observables $\delta g_{L,R}/g_{L,R}$, which appear when we use different localization of bulk fermions in order to solve the flavour problem. This mechanism would also generate flavor-changing neutral currents (FCNC) and CP violating effective operators. However, as the issue of fermion flavor will not be covered in this review, we will concentrate in the following sections on constraints from universal corrections. In section 4 we will consider the case of AdS$_5$, the original RS proposal. We will deduce the constrains on the Higgs profile in order to solve the hierarchy problem and the constraints imposed on the mass of the first KK excitation of gauge bosons $m_{KK}$ from electroweak precision tests (EWPT). In particular we will find that, depending on the Higgs localization, EWPT impose lower bounds $m_{KK}\gtrsim 7 \textrm{ TeV}-13$ TeV, which are outside the LHC reach and would recreate a little hierarchy problem. This problem has motivated extensions of the RS model in order to overcome those bounds. Two possibilities have been reviewed here. 
The first possibility, considered in section 5, consists in introducing a dilatonic scalar such that the warping of the fifth dimension is strongly modified near the IR brane while it behaves as AdS$_5$ near the UV region. In fact we are using soft-wall metrics such that the naked singularity is regularized by the IR brane. The closer the IR brane to the naked singularity the stronger the effect. The hierarchy problem imposes constrains on the Higgs profile which are stronger than those in the RS case. It turns out that there is a strong renormalization wave equation for the Higgs field such that there is an extra suppression on electroweak observables. This renormalization effect effectively depends on the departure of the IR metric from AdS$_5$. Electroweak constraints then provide mild bounds as $m_{KK}\gtrsim 1\text{ TeV}-3$ TeV depending on the values of the parameters defining the metric. 
The second way out to the problem of strong electroweak constraints on the pure RS model has been reviewed in section 6. It consists in adding a gauge symmetry in the bulk which contains the custodial symmetry group $SU(2)_R$ and thus guarantees that the SM tree-level condition $T=0$ is respected, thanks to the cancellations induced by the extra states required to accommodate matter in $SU(2)_R$ multiplets, while there is a tree-level contribution to the $S$ parameter. In this case the one-loop contribution to the $T$ parameter coming from the top quark and its $SU(2)_R$ partners should be considered. Finally EWPT impose the condition $m_{KK}\gtrsim 3$ TeV. Moreover we also review the case where one embeds the custodial group $SU(2)_R$ into a larger group such that the Higgs degrees of freedom can be identified with the fifth components of the 5D gauge fields along the broken generators: these models are dubbed Gauge-Higgs Unification models. In Gauge Higgs Unification the Higgs potential and the Higgs mass can be obtained dynamically by the Coleman-Weinberg mechanism. Different models then essentially differ by the bulk group $\mathcal G$ containing $SU(2)_R$ and by the subgroup surviving at the IR brane and they are thus dubbed with different names, e.g.~minimal composite Higgs model for $\mathcal G=SO(5)$. 
Finally section~7 contains our conclusions which essentially will depend on future experimental results on the LHC13-14 running where Higgs data on production strengths will be delivered with more accuracy. In particular they should clarify whether there are any departures with respect to the Standard Model predictions in the different channels.

\section{Standard Model in Warped Extra Dimensions}
We will now consider the Standard Model propagating in a 5D space with an arbitrary metric $A(y)$ such that in proper coordinates the line interval is: 
$
ds^2=e^{-2A(y)}\eta_{\mu\nu}dx^\mu dx^\nu+dy^2
$
where $\eta_{\mu\nu}=\textrm{diag}(-1,1,1,1)$. This anstaz is the most general one consistent with Minkowski space-time in 4D. We define the 5D $SU(2)_L\times U(1)_Y$ gauge bosons as $W^i_M(x,y)$, $B_M(x,y)$ where $i=1,2,3$ and $M=\mu,5$, and the SM Higgs as
\begin{equation}
H(x,y)=\frac{1}{\sqrt 2}e^{i \chi(x,y)} \left(\begin{array}{c}0\\h(y)+\xi(x,y)
\end{array}\right)
\label{Higgs}
\end{equation}
where the matrix $\chi(x,y)$ contains the three 5D SM fields $\vec\chi(x,y)$. The Higgs background $h(y)$ as well as the metric $A(y)$ are for the moment arbitrary functions and they will be specified later on. We will consider the 5D action for the gauge fields and the Higgs field $H$ as:
\begin{eqnarray}
S_5&=&\int d^4x dy\sqrt{-g}\left(-\frac{1}{4} \vec W^{2}_{MN}-\frac{1}{4}B_{MN}^2-|D_M H|^2-V(H)
\right)\nonumber\\
&-&2\sum_{\alpha}\int d^4x dy \sqrt{-\bar g_\alpha}\,(-1)^\alpha\,\lambda^\alpha(H)\delta(y-y_\alpha)
\label{5Daction}
\end{eqnarray}
where $V$ is the 5D potential, $\lambda^\alpha$ ($\alpha=0,1$) the 4D brane potentials, $\bar g_\alpha$ the induced metric at the brane located at $y_\alpha$, and we will choose conventionally $y_0=0$. The bulk $V$ and brane $\lambda^\alpha$ potentials will depend in general, not only on the Higgs $H$ but also on possible stabilizing (dilatonic) fields $\phi$ in the theory. 
From here on we will assume that $V(H)$ is quadratic in $H$. 
Electroweak symmetry breaking will be triggered on the IR brane. We thus choose the brane potentials as
\begin{equation}
\lambda^0(H)=M_0 |H|^2
\,,\qquad
-\lambda^1(H)=-M_1 |H|^2+\gamma |H|^4\,.
\label{boundpot}
\end{equation}
The effect of the warp factor can be encoded in the parameter $\rho$ defined as
\begin{equation}
\rho=ke^{-A(y_1)}
\label{rho}
\end{equation}
and in order to solve the hierarchy problem $A(y_1)$ has to be chosen such that for a Planckian value of $k$, $\rho$ is at the TeV scale: i.e.~$A(y_1)\simeq 35$.

One can then construct the 4D effective theory out of (\ref{5Daction}) by making the KK-mode expansion
$
A_\mu(x,y)\equiv \sum_nA_\mu^n(x)\cdot f_A^n(y)/\sqrt{y_1}
$
where $A_\mu=A^\gamma_\mu,\,Z_\mu,\,W^{\pm}_\mu$. The functions $f_A$ satisfy the equations of motion (EOM)
$
m_{f_A}^2 f_A+(e^{-2A}f^{\prime}_A)'-M_A^2 f_A=0,
$
they are normalized as
$
\int_0^{y_1}f_A^2(y)dy=y_1
$,
and satisfy the Neumann boundary conditions (BC)
$
\left. f^{\prime}_A\right|_{y=0,y_1}=0.
$
We have defined the 5D $y$-dependent gauge boson masses as
$
M_W(y)=\frac{g_5}{2} h(y)e^{-A(y)},\,
M_Z(y)=\frac{1}{c_W} M_W(y),\,
M_\gamma(y)
\equiv 0
$
where $c_W={g_5}/{\sqrt{g_5^2+g_5'^2}}$, and $g_5\equiv g\sqrt{y_1}$, $g'_5\equiv g'\sqrt{y_1}$, are the 5D $SU(2)_L$ and $U(1)_Y$ couplings respectively. Only the lightest mass eigenvalue will be significantly affected by the breaking so we simplify our notation by defining
$
m_A=m_{f_A^{0}}\,,%\qquad f_A=f_A^{0}
$
for the zero modes and
$
m_{n}=m_{f_A^{n}},\, f^{n}=f^{n}_A\,,
$
for the higher modes ($n\geq 1$). An approximated expression is given by~\cite{CGQ1,CGQ2,CGQ3,CGQ4}
\begin{equation}
m_A^2\simeq m_{A,0}^2\equiv\frac{1}{y_1}\int_0^{y_1}dy\, M_A^2(y)\,.
\label{gaugezero}
\end{equation}

The background $h(y)$ is determined from the bulk EOM and BC
\begin{eqnarray}
&h''(y)-4A'h'(y)-\partial V/\partial h=0\nonumber\\
&h'(y_\alpha)={\displaystyle \left.\partial\lambda^\alpha/\partial h\right|_{y_\alpha}},
\label{EOMHiggs}
\end{eqnarray}
while for the Higgs fluctuations, the bulk EOM and BC are
\begin{eqnarray}
&\xi''(y)-4A'\xi'(y)-\left(\partial^2 V/\partial h^2\right)\xi(y)+m_H^2 e^{2A}\xi(y)=0\nonumber \\
&\xi'(y_\alpha)/\xi(y_\alpha)=\left.\partial^2 \lambda^\alpha(h)/\partial h^2\right|_{y_\alpha}.
\label{EOMHiggsfluc}
\end{eqnarray}
With our choice (\ref{boundpot}) of boundary potential the UV boundary conditions for the background and fluctuations are the same, and  for a quadratic bulk Higgs potential the Higgs wave function $\xi(y)$ for $m_H=0$ ($n=0$) is thus proportional to $h(y)$. For small Higgs mass this will still be a good approximation to the exact wave function. This means that the 5D VEV will be carried almost entirely by the zero mode. 

We can then write an effective theory by the decomposition $H(x,y)=\sqrt k\,\mathcal H(x)h(y)/h(y_1)$ and calculate the effective Lagrangian for the mode $\mathcal H(x)$. In fact the SM Lagrangian is
$
\mathcal L_{SM}=-\left|\mathcal D_\mu H_{SM}\right|^2+\mu^2|H_{SM}|^2-\lambda |H_{SM}|^4 \,,
$
where the SM parameters $\mu^2$ and $\lambda$ are related to 5D quantities by~\cite{CGQ2}
\begin{eqnarray}
%H_{SM}(x)&=& \sqrt{Z}e^{-A(y_1)} \mathcal H(x)\,,\\
\mu^2&=&(kZ)^{-1}\left(M_1-\frac{h'(y_1)}{h(y_1)}  \right)\; \rho^2\,,\quad
\lambda=\frac{\gamma k^2}{Z^2}\label{gammaSM}\,,\nonumber\\
 Z&\equiv&
k\int_0^{y_1}dy\frac{h^2(y)}{h^2(y_1)}e^{-2A(y)+2A(y_1)}
\label{muSM}
\end{eqnarray}
and 
\begin{equation}
H_{SM}(x)\equiv \sqrt{Z}e^{-A(y_1)} \mathcal H(x)\,.
\label{rescaling}
\end{equation}
 The usual SM expressions for the Higgs VEV $v^2=\mu^2/\lambda$ and Higgs mass $m_H^2=2\mu^2$ follow straightforwardly. The quantity $Z$ is an additional wave function renormalization depending on both the gravitational and Higgs backgrounds. As we will see in the next section (see also Refs.~\cite{CGQ1} and \cite{CGQ2}), a sizable $Z$ suppresses the $T$ ($S$) parameter by two (one) powers of $Z$, which leads to a corresponding reduction in the value of $\rho$ from EWPT. Moreover we see here that the parameter $\mu$, or equivalently the Higgs mass, is further reduced with respect to $\rho$ by a factor $1/\sqrt{Z}$ which in turn reduces the required amount of fine-tuning in the factor $M_1-h'(y_1)/h(y_1)$.

\section{Electroweak precision observables}
 There are three experimental input parameters, usually referred to as $S$, $T$ and $U$ parameters~\cite{Peskin:1991sw}, which appear in theories where fermions couple in a universal way: they are called oblique parameters. In models with a gap between the electroweak and new physics scales the $U$ parameter is expected to be small since it corresponds to a dimension eight operator. On the other hand, there are other dimension six operators unrelated to electroweak breaking which in some models can have sizable coefficients.
It has thus been suggested to instead consider the set $T$, $S$, $Y$ and $W$  as a more adequate basis for models of new physics~\cite{Barbieri:2004qk}. They are defined as%
\begin{eqnarray}
\alpha T&=&
  m_W^{-2}\left[c_W^2 \Pi_{Z}(0)-\Pi_{W}(0)\right]\,,\quad
\alpha S=
4s_W^2c_W^2\left[\Pi'_{Z}(0)-\Pi_{\gamma}'(0)\right]\,,\nonumber\\
2m_W^{-2}Y&=&
s_W^2\Pi_{Z}''(0)+c_W^2\Pi_{\gamma}''(0)\,,\quad
2m_W^{-2}W=
c_W^2\Pi_{Z}''(0)+s_W^2\Pi_{\gamma}''(0)\,,
\label{ST}
\end{eqnarray}
where $\alpha$ is the electromagnetic gauge coupling defined at the $Z$-pole mass. 

In theories with a Higgs mode $H$ of mass $m_H\ll m_{\rm KK}$ one can relate $T$, $S$, $Y$ and $W$ to the coefficients of the dimension six operators
\begin{equation}
|H^\dagger D_\mu H|^2\,,\qquad
H^\dagger W_{\mu\nu}HB^{\mu\nu}\,,\qquad
(\partial_\rho B_{\mu\nu})^2\,,\qquad (D_\rho W_{\mu\nu})^2
\label{ops}
\end{equation}
in the effective low energy Lagrangian respectively. These coefficients can be obtained in the effective theory by integrating out the KK-modes from where one can write the general expression~\cite{CGQ2} 
\begin{eqnarray}
\alpha T&=&s_W^2m_Z^2y_1\int e^{2A}(\Omega_f-\Omega_h)^2\,,\nonumber\\
\alpha S&=&8s_W^2c_W^2 m_Z^2 y_1\int e^{2A}\left(\Omega_f-\frac{y}{y_1}\right)(\Omega_f-\Omega_h)\,,\nonumber\\
W&=&Y=c_W^2m_Z^{2}\,y_1
\int e^{2A}\left(\Omega_f-\frac{y}{y_1}\right)^2
 \label{STYW}
\end{eqnarray}
where the function $\Omega_f$, determined by the localization properties of fermion $f$ with a 5D Dirac mass equal to $m_D^f=c_f A'(y)$, is given by
\begin{equation}
 \Omega_f(y)=\frac{\int_0^{y}e^{(1-2c_f)A}}{ \int_0^{y_1}e^{(1-2c_f)A}}
\end{equation}
and we are assuming for the moment universal localization of all fermions. Similarly the localization properties of the Higgs field are determined by the function $\Omega_h$ given by
\begin{equation}
\Omega_h(y)=\frac{\int_0^y h^2(y')e^{-2A(y')}}{\int_0^{y_1}h^2(y')e^{-2A(y')}} \,.
\label{Omega}
\end{equation}

In the case of UV-localized fermions $(c_f\to\infty)$ we can see that $\Omega_f(y)=1$. This is the usually considered case as first and second generation fermions, which are the relevant ones for most of the precision observables, are light and then $c_f>1/2$ for them. In the case of flat fermions, i.e.~$c_f=1/2$ we have that $\Omega_f=y/y_1$ so that $S=W=Y=0$ but still $T$ is non-vanishing and it has to be taken into account for phenomenological purposes. Finally for the case of IR localized fermions ($c_f<1/2$) $\Omega_f(y)$ is localized towards the IR brane. As also the Higgs is localized towards the IR brane in order to solve the hierarchy problem (as we will see later on) it turns out that also $\Omega_h(y)$ is localized towards the IR brane and thus the difference $\Omega_f-\Omega_h$, along with the $T$ parameter, tends to be small, but still the contribution from the other parameters $S$, $W$ and $Y$ can give very strong phenomenological constraints. 
 
If fermion localization is not universal, i.e.~if the constants $c_f$ are not equal for all fermions, as it is usually the case when one wants to provide a theory of fermion flavor from fermion localization, then non-oblique corrections are also generated. In particular the most important of those non-oblique operators give corrections to the $Zb\bar b$ coupling. The calculation was done in Ref.~\cite{CGQ5} with the result
\begin{eqnarray}
\delta g_{b_{L,R}}&=&%\delta g_{d_{L,R}}
\frac{g^{SM}_{d_{L,R}}}{2} \left(\alpha T+\frac{Y}{c_w^2}\right)
+\frac{1}{3}\,\frac{1}{c_w^2-s_w^2}\left(\frac{\alpha S}{4}-c_w^2 s_w^2\,\alpha T-s_w^2\,Y\right)
+\delta\tilde g_{b_{L,R}}\,,\nonumber\\
\delta\tilde g_{b_L}&=&\left(-g^{SM}_{d_{L}}\, m_Z^2\, 
\hat\alpha_{h,d^i_{L}} \delta_{i\ell}+\frac{v^2}{4}\beta^{d_L}_{i\ell}\right)
V^{3i}_{d_L}V_{d_L}^{*3 \ell}\,,\nonumber\\
\delta \tilde g_{b_R}&=&\left(-g^{SM}_{d_{R}}\, m_Z^2\, 
\hat\alpha_{h,d^i_{R}} \delta_{i\ell}-\frac{v^2}{4}\beta^{d_R}_{i\ell}\right)
V^{3i}_{d_R}V_{d_R}^{*3 \ell}\,,
\label{deltagoverg}
\end{eqnarray}
where $g_{q_L}^{SM}=T^3_{q}-Q^{em}_q\,s_w^2$ and $g^{SM}_{q_R}=-Q^{em}_qs_w^2$  and the integrals $\hat\alpha_{hf}$ and $\beta_{ij}^\psi$ are
\begin{eqnarray}
\hat \alpha_{hf}&=&y_1\int e^{2A}\left(\Omega_h-\frac{y}{y_1}\right)(\Omega_f-1)\nonumber\,,\\
\beta^{d_L}_{i\ell}&=&
\sum_jY_{ij}^{d}Y_{\ell j}^{d*}
\int_0^{y_1}
e^{2A}
\,
(\Omega'_{d_R^j})^{-1}
(\Gamma^d_{\ell j}-\Omega_{d^j_R})(\Gamma^d_{ij}-\Omega_{d_R^j})\,,\nonumber\\
\beta^{d_R}_{i\ell}&=&
\sum_jY_{ji}^{d^*}Y_{j\ell}^{d}
\int_0^{y_1}
e^{2A}
\,
(\Omega'_{d_L^j})^{-1}
(\Gamma^d_{j\ell}-\Omega_{d^j_L})(\Gamma^d_{ji}-\Omega_{d_L^j})\,,\nonumber\\
\Gamma^d_{ij}(y)&=&\frac{\int_0^{y}h\,e^{-(c_{Q_L^i}+c_{d_R^j})A}}{\int_0^{y_1}h\,e^{-(c_{Q_L^i}+c_{d_R^j})A}}\,,
\end{eqnarray}

As the bounds from the $Zb\bar b$ coupling depend, to a large extent, on the different localization of bulk quarks in order to reproduce the texture of quark masses and mixing angles, a topic which will not be covered by this article, we will concentrate on the case of oblique corrections and we will assume that approximately first and second generation quarks are localized towards the UV brane, in which case $\Omega_f=1$ in Eq.~(\ref{ST}). We can then readily compute the electroweak observables as
\begin{eqnarray}
\alpha T&=&s_W^2 m_Z^2\, \frac{I_2}{\rho^2}\, \frac{ky_1}{Z^2}\,\,,\nonumber\\
\alpha S&=&8 s^2_W c^2_W m_Z^2\, \frac{I_1}{\rho^2} \,\frac{1}{Z} \,,\nonumber\\
Y&=&W=c_W^2 m_Z^2\, \frac{I_0}{\rho^2}\,\frac{1}{ky_1}\,,\nonumber\\
\label{STYWfinal}
\end{eqnarray}
where 
\begin{equation}
I_n=k^{3}\int_0^{y_1}dy (y_1-y)^{2-n}u^n(y)e^{2A(y)-2A(y_1)},\quad u(y)=\int_y^{y_1}dy'\frac{h^2(y')}{h^2(y_1)} e^{-2A(y')+2A(y_1)}.
\label{notacion}
\end{equation}
The dimensionless integrals $I_n$ are expected to be of the same order. In particular one expects $I_n/\rho^2=\mathcal O(1/m^2_{KK})$. 
We see that $T$ is enhanced by a volume factor and thus it is expected to be the leading observable, while $S$ carries no power of the volume and it is thus the next to leading one. On the other hand the $W$ and $Y$ parameters are suppressed with one (two) additional volume factor(s) as compared to $S$ ($T$).

\section{AdS$_5$}

The paradigm of warped space was first introduced by Randall and Sundrum~\cite{Randall:1999ee} and it consists in a 5D AdS space, i.e.~with a metric given by $A(y)=ky$, with two branes localized at $y=0$ (the UV brane) and at $y=y_1$ (the IR brane).  In this case the wave function of KK gauge bosons is obtained analytically (neglecting the small correction from EWSB) as
\begin{equation}
f_A^n=N_ne^{ky}\left\{J_1\left(\frac{m_n}{k}e^{ky}\right)+a_n Y_1\left(\frac{m_n}{k}e^{ky}\right)\right\}
\end{equation}
 where $J_1$ and $Y_1$ are Bessel functions and the constant $a_n$ and the mass eigenvalues $m_n$ come from imposing the UV and IR boundary conditions. $k$ is the AdS curvature and it is warped down to the TeV scale by $\rho\equiv e^{-ky_1}k$ while the mass of the first KK mode is related to $\rho$ by the relation: $m_{KK}=c_K \rho$ with $c_K\simeq  2.44$.
 
\subsection{The hierarchy problem}
The original solution to the hierarchy problem comes for a Higgs field fully localized in the IR brane~\cite{Randall:1999ee}. In this case we can write the Higgs action as
\begin{equation}
S_H=\int d^4x \sqrt{-\bar g_1}\left\{  -\bar g_1^{\mu\nu}D_\mu H^\dagger D_\nu H-\lambda(|H|^2-v_0)^2\right\}
\end{equation}
where $\bar g_{1\,\mu\nu}\equiv e^{-2ky_1}\eta_{\mu\nu}$ is the induced metric at the IR brane. After making the rescaling $H\to e^{ky_1}H$ the action can be written as
\begin{equation}
S_H=\int d^4x \left\{  -|D_\mu H|^2-\lambda(|H|^2-e^{-ky_1}v_0)^2\right\}
\end{equation}
from where it can be seen that a Planckian value of $v_0$ is brought to the electroweak value $v$ as $v\equiv e^{-ky_1}v_0$. In fact this result is completely general: any mass parameter in the IR brane $m_0$ get rescaled in the same way as $m\equiv e^{-ky_1}m_0$. This is the original warped solution to the hierarchy problem.

Of course having a Higgs localized on the IR brane is not the only solution to the hierarchy problem as we will see next. In fact, for a quadratic bulk Higgs potential as $V(H)=k^2 a(a-4)|H|^2$, to determine whether the hierarchy problem is successfully solved by a profile of the form $h(y)\sim e^{aky}$ with small $a\simeq \mathcal O(1)$ let us write the solution of our Higgs profile subject to a generic electroweak symmetry preserving UV-boundary condition (\ref{EOMHiggs}) as
 \begin{equation}
h(y)=h_0\left(\frac{M_0/k+a-4}{2(a-2)}e^{a k y}
-\frac{M_0/k-a}{2(a-2)}e^{(4-a) k y}\right)
\label{RSsoln}
\end{equation}
For $a>2$ no fine-tuning is necessary in order to keep only the first term, since near the IR brane (where EWSB occurs) the second term is always irrelevant. On the contrary, for $a<2$, the second term would be dominating and one needs to fine-tune $M_0/k\simeq a$ in order to maintain the solution $h(y)\simeq e^{aky}$. This fact has a simple holographic interpretation: since $\dim(\mathcal O_H)=a$ the hierarchy problem is solved by the Higgs compositeness for $a>2$, but not for $a<2$ (see Ref.~\cite{Luty:2004ye} for a more detailed discussion). One can now quantify using 5D techniques the required amount of fine-tuning for $a<2$ by imposing $M_0=k a+\Delta M_0$ and assuming that the second component in (\ref{RSsoln}) is similar to the first component at the IR brane, we can easily quantify the amount of required fine-tuning as
\begin{equation}
\frac{\Delta M_0}{M_0}\simeq 2\frac{2-a}{a}e^{-2(2-a)ky_1}.
\label{FTRS}
\end{equation}
In fact Eq.~(\ref{FTRS}) yields, for $a=1$, the same amount of fine-tuning which is required in the Standard Model, while it decreases linearly in the exponent along the interval $1<a<2$.

In this theory the function $Z$ in Eq.~(\ref{muSM}) is given by $Z=\frac{1}{2(a-1)}$ and the value of the Higgs mass from (\ref{muSM}) is given by
\begin{equation}
m_H^2=\frac{4(a-1)}{c_K^2}(M_1/k-a)m_{KK}^2.
\end{equation}
Then the required amount of fine-tuning for obtaining a light Higgs boson can be quantified by imposing $M_1=ak+\Delta M_1$ as
\begin{equation}
\frac{\Delta M_1}{M_1}=\frac{c_K^2}{4a(a-1)} \frac{m_H^2}{m_{KK}^2}.
\end{equation}
This would correspond to a $\sim$ 1\% fine-tuning for $m_{KK}\sim 1$ TeV. However, as we will see in the next section, electroweak observables put very strong lower bounds on the value of $m_{KK}$, which translate into a more severe tuning for a Higgs with a mass $m_H=126$ GeV as experimentally observed at the LHC. For instance for $m_{KK}\sim 10$ TeV the required tuning is $\sim 10^{-4}$: this is the typical tuning in the RS model and it would create a little hierarchy problem associated with the KK masses.

\subsection{Electroweak constraints}
For the AdS$_5$ metric it is very simple to compute analytically the $T$, $S$, $W$ and $Y$ parameters from Eq.~(\ref{STYW}). In fact they are given by~\cite{CGQ1}
\begin{eqnarray}
\alpha T&=&s_W^2 c_K^2 ky_1 \frac{(a-1)^2}{a(2a-1)}\frac{m_Z^2}{m_{KK}^2}\nonumber\\
\alpha S&=& 2 s_W^2 c_W^2 c_K^2 \frac{a^2-1}{a^2}\frac{m_Z^2}{m_{KK}^2}\nonumber\\
W&=&Y=c_W^2 c_K^2 \frac{1}{4ky_1}\frac{m_Z^2}{m_{KK}^2}
\label{STYW-RS}
\end{eqnarray}
from where we can see that, as foreseen for general metrics, the $T$ parameter is enhanced by the compactification volume $(ky_1)$ and it is thus the most constraining one out of the oblique parameters. On the other hand $S$ does not depend on the volume and it is the next to leading parameter, while $W$ and $Y$ are suppressed by the compactification volume and they are thus usually negligible in phenomenological applications.  In the following we will just consider the leading $T$ and $S$ parameters.
\vspace{0.5cm}
\begin{figure}[htb]
\centerline{\psfig{file=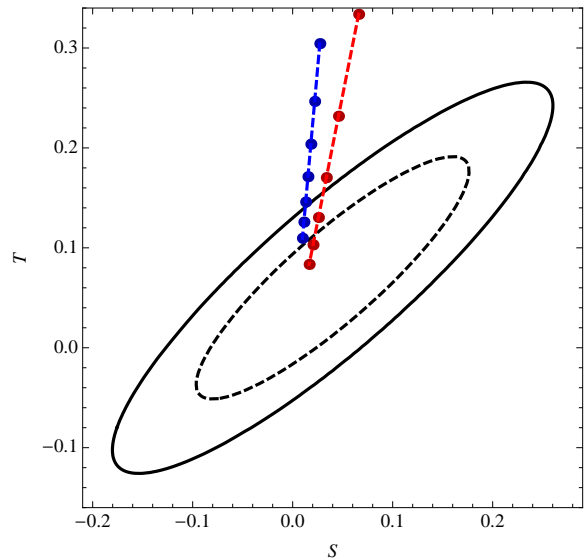,width=3.0in}}
\vspace*{8pt}
\caption{Outer (inner) ellipse is the allowed region at the 95\% CL (68\% CL). Right (red) dots are predictions for $a=2$ and $5 \text{ TeV}\leq m_{KK}\leq 10$ TeV, $\Delta m_{KK}= 1$ TeV. Left (blue) dots are predictions for an IR localized Higgs and $9 \text{ TeV}\leq m_{KK}\leq 15$ TeV, $\Delta m_{KK}= 1$ TeV. In all cases the largest value of $m_{KK}$ corresponds to the most inner dot inside the ellipses.
\protect\label{fig1}}
\end{figure}

Comparison of oblique parameters with experimental data provide in general lower bounds on the KK masses $m_{KK}$.  In Fig.~\ref{fig1} we show the 95\% CL (68\% CL) allowed region, for a Higgs mass $m_H\sim 126$ GeV, which corresponds to the region inside 
the outer (inner) ellipse~\cite{PDG}. The dots and interpolating lines correspond to the predictions of the RS model for $a=2$ [right (red dots) dashed line] and for the limit $a\to\infty$ [left (blue dots) dashed line]. The values for the $a=2$ (red) dots correspond to $5 \text{ TeV}\leq m_{KK}\leq 10$ TeV where the dots are spaced by $\Delta m_{KK}=1$ TeV, and $m_{KK}\simeq 7$ TeV corresponds to the dot touching the 95\% CL ellipse while $m_{KK}=5$ TeV corresponds to the most outer dot. Similarly the line corresponding to an IR localized Higgs corresponds to $9 \text{ TeV}\leq m_{KK}\leq 15$ TeV where the dots are spaced by $\Delta m_{KK}=1$ TeV and $m_{KK}\simeq 13$ TeV is the dot touching the 95\% CL ellipse while $m_{KK}=9$ TeV is the furthest dot on the blue line. 

Therefore from the results of Fig.~\ref{fig1} we conclude that, for an IR localized Higgs, compatibility with EWPT provides the lower bound $m_{KK}\gtrsim 13$ TeV at the 95\% CL, while the bound improves when we delocalize the Higgs from the IR brane as much as possible. The solution of the hierarchy problem then implies that $a\geq 2$ and we get the global bound $m_{KK}\gtrsim 7$ TeV at the 95\% CL. 

The model described in this section has of course the capability of providing a theory of quark and lepton  flavor by adjusting the constants $c_f$ to fit all fermion masses and mixing angles contained in the CKM~\cite{CGQ5} and the PMNS~\cite{vonGersdorff:2012tt} matrices. This process is not for free because it originates extra FCNC and CP violating operators which further constrain the model parameters. These developments are outside the scope of this review article and will be covered in Ref.~\cite{Gero}.

However it goes without saying that the RS model we have just presented is not phenomenologically appealing as:
\begin{itemize}
\item
KK modes are not accessible to LHC.
\item
It recreates a little hierarchy problem.
\end{itemize} 

The ways of relaxing the strong constraints from the electroweak observables consist obviously in modifying the model as will be explained in the next sections. We will see that there are mainly two options to overcome this problem:
\begin{itemize}
\item
Either one can deform the metric in the IR region in such a way that it is still AdS$_5$ asymptotically near the UV brane (AAdS$_5$). As we will see the deformation can create a naked singularity of the metric. By localizing the IR brane close enough to the singularity its effect will damp the impact of the $T$-observable on the lower KK mass bounds. In this kind of background we do not need to add extra matter and/or gauge bosons and we can just propagate the Standard Model content in the bulk.
\item
Or another option is to keep the AdS$_5$ geometry everywhere but to extend the gauge group and matter content in the bulk. In particular we will see that introducing a local $SU(2)_R$ group in the bulk (which holographically corresponds to a global custodial symmetry group in 4D) the $T$-observable can be protected from radiative corrections and again the bounds on KK mass observables are mitigated. We will call these Custodial Models.
\end{itemize} 

Let us finally mention another way of relaxing constraints from electroweak observables (although it will not be covered in this review). It consists~\cite{Davoudiasl:2002ua,Carena:2002dz} in introducing brane-localized kinetic terms for gauge bosons and/or fermions. These terms will be generated anyway from radiative corrections at the orbifold fixed points (where branes are localized). The shortcoming of this approach is that the required strengths for overcoming the constraints from electroweak observables are much larger that the natural values generated by radiative corrections.

\section{Standard Model in singular backgrounds}
Let us assume the 5D background metric is such that there is a naked singularity at the point $y=y_s$ and that we locate the IR brane at $y=y_1<y_s$, a short distance away from the singularity, i.e.~$y_s=y_1+\Delta$ with $k\Delta\ll k y_s$. In principle this can be done by introducing a stabilizing diatonic field $\phi$ with potentials at the UV and IR branes fixing appropriates values $\phi_0\equiv\phi(0)$ and $\phi_1\equiv\phi(y_1)$. In this case as the metric has a singularity at $y=y_s$ the integral defining the Higgs wave function renormalization $Z$ in Eq.~(\ref{muSM}) is dominated by the region near $y_1$ and it is thus enhanced with respect to the RS value. Constructions based on singular backgrounds are usually dubbed soft-walls~\cite{Gubser,AdS/QCD,Batell1,Falkowski1,Batell2}.

One particularly simple construction was proposed in Ref.~\cite{Cabrer:2009we} (see also Ref.~\cite{Brouzakis:2013gda}) where the metric and dilaton profiles were given by
\begin{eqnarray}
\phi(y)&=&-\frac{\sqrt{6}}{\nu}\log[\nu^2 b k(y_s-y)]
\label{phi}
\ ,
\\
A(y)&=&ky-\frac{1}{\nu^2}\log\left(1-\frac{y}{y_s}\right)
\ ,
\label{A}
\end{eqnarray}
where $\nu$ is a real parameter and we are using the normalization $A(0) =  0$. As we can see from (\ref{A}) near the UV brane ($y\sim 0$) the metric coincides with that of AdS$_5$ while when $y\to y_s$ there is a large departure from the AdS$_5$ metric: this is the typical behaviour of AAdS$_5$.

\begin{figure}[htb]
\centerline{\psfig{file=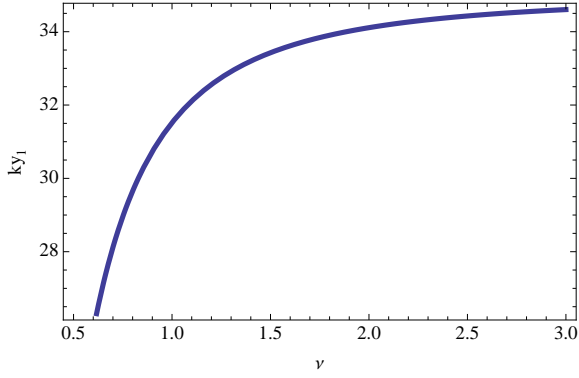,width=3.0in}}
\vspace*{8pt}
\caption{$ky_1$ as a function of $\nu$ for $A(y_1)=35$ and $k\Delta=1$.
\protect\label{fig2}}
\end{figure}

As we said we are assuming that the brane dynamics $\lambda^\alpha(\phi)$ fixes the values of the field \mbox{$\phi=(\phi_0,\,\phi_1)$} on the UV and IR branes respectively. The interbrane distance, $y_1$, as well as the location of the singularity, $y_s\equiv y_1+\Delta$, are related to the values of the field $\phi_\alpha$ at the branes by the following simple expressions:
\begin{equation}
ky_1=\frac{1}{\nu^2}\left[e^{-\nu \phi_0/\sqrt{6}}-e^{-\nu \phi_1/\sqrt{6}}  \right], \quad k\Delta=\frac{1}{\nu^2}e^{-\nu \phi_1/\sqrt{6}}
\ ,
\end{equation}
which shows that the required large hierarchy can naturally be 
fixed with values of the fields $|\phi_\alpha|=\mathcal O(1)$ and $\phi_1\gtrsim \phi_0$, $\phi_0<0$. Note that due to its exponential dependence on $\phi_1$, $k\Delta$ can be small or, in other words, the IR brane can naturally be located very close to the singularity. One interesting feature of this model is that the value of the volume $ky_1$ required to solve the hierarchy problem is smaller than in the model with AdS$_5$ as it is shown in Fig.~\ref{fig2} where we plot $y_1$ as a function of $\nu$ for $k\Delta=1$ (which we will be using in the rest of this paper) and $A(y_1)=35$.
We see that for values of $\nu<1$ there can be a considerable reduction on the volume $ky_1$. On the contrary for large values of $\nu$ the RS-limit is recovered and consequently the volume $ky_1\simeq  35$.

The EOM, e.g.~for the gauge boson KK modes, cannot be solved analytically in such a complicated background so that all the calculations have to be performed numerically. A useful fit relating the parameter $\rho$ with the mass of the first KK excitation $m_{KK}$ is given by Ref.~\cite{CGQ1}
\begin{equation}
m_{KK}\equiv c_K\, \rho\simeq \left[2.44+\frac{1.71}{\nu^2}\right]\rho\ .
\label{fit}
\end{equation}

\subsection{The hierarchy problem}
The bulk Higgs potential is written as
\begin{equation}
V(H)=k^2[a(a-4)-4 a b e^{\nu\phi/\sqrt{6}} ]|H|^2
\end{equation}
and the coefficients of the two operators $|H|^2$ and $e^{\nu\phi/\sqrt{6}}|H|^2$ can be considered as independent parameters. However since the parameter $b$ can be traded by a global shift in the value of the $\phi$ field, or in particular by a shift in its value at the UV brane $\phi_0$, for simplicity we will fix its value to $b=1$ hereafter. We can now write the general solution to Eq.~(\ref{EOMHiggs}) as
\begin{equation}
h(y)=e^{aky}\left(c_1+c_2\int^y e^{-2(a-2) k y'}\left(1-\frac{y'}{y_s}\right)^{-\frac{4}{\nu^2}}\right)\,.
\end{equation}
The two integration constants $c_i$ are fixed from the boundary conditions in (\ref{EOMHiggs}) derived from the boundary potentials $\lambda^\alpha$ given in Eq.~(\ref{boundpot}).

\begin{figure}[htb]
\centerline{\psfig{file=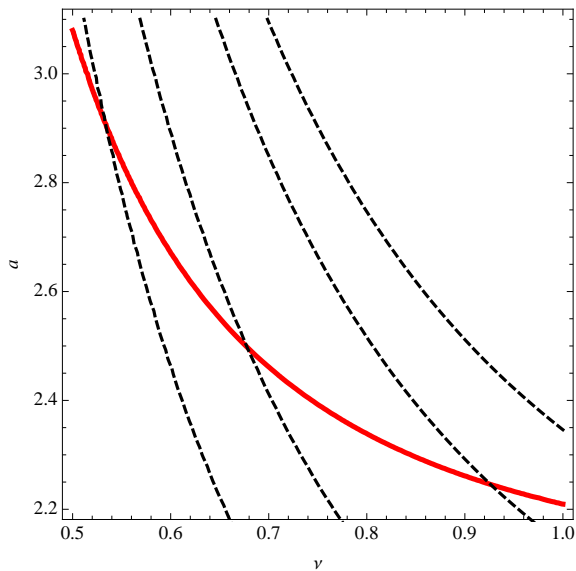,width=3.0in}}
\vspace*{8pt}
\caption{Solid (red) line is the contour plot for $\delta=1$ in the $(\nu,a)$ plane. Dashed (black) lines are contour plots of constant values of $Z=0.75$ (right line) 1, 2 and 5 (left line).
\protect\label{fig3}}
\end{figure}

In order to determine whether the hierarchy problem is successfully solved by a profile of the form $h(y)\sim e^{aky}$ with small $a$ let us write the solution of our Higgs profile subject to a generic electroweak symmetry preserving UV-boundary condition $h'(0)=M_0h(0)$ as
\begin{equation}
h(y)=h_0e^{a k y}\left(1+[M_0-a k]\int_0^y e^{4A(y')-2a k y'}\right)\,.
\label{fullsoln}
\end{equation}
As in the AdS$_5$ case, for $a<a_{\rm min}$~\footnote{Note that while $a_{min}=2$ for RS, $a_{\rm min}>2$ for the singular metric (\ref{A}) as we will see next.} the solution $h(y)\sim e^{a k y}$  will  be fine-tuned due to the exponential enhancement in the integrand. We can rewrite the solution as
\begin{equation}
h(y)
=h_0 e^{aky} \bigg[
1 + (M_0/k -a) \left[ F(y) - F(0) \right]
\bigg]
,
\label{eq:hyF}
\end{equation}
where the function $F(y)$ is given by
\begin{equation}
F(y) =  e^{-2(a-2) k y_s}k y_s \left[ -2(a-2) k y_s \right]^{-1 + 4/\nu^2} \Gamma \left[ 1 - \frac{4}{\nu^2} , -2(a-2) k( y_s - y) \right]
\end{equation}
Note that $F(y)$ is defined as a complex function but its imaginary parts cancels in \eqref{eq:hyF} leading to a real function. One should view $F(y)$ as the generalization of $F_{\rm RS}(y)=e^{-2(a-2)ky}$ in the RS case. In order to keep the exponential solution without the need of any fine-tuning we must require the function $F(y)$ to be small in absolute value. Since $|F(y)|$ is a monotonically increasing function of $y$ it will be enough to inspect $\delta\equiv |F(y_1)|$ which will be a measure of the fine-tuning required in $(M_0/k - a)$ in order to keep the exponential solution. In particular the absence of fine-tuning requires roughly $\delta \lesssim \mathcal O(1)$. 
The contour plot for $\delta=1$ in the plane ($\nu,a$) with $k\Delta=1$ is plotted in the solid (red) line of Fig.~\ref{fig3}~\footnote{In the considered region of the parameter space we can see that the value $a=3.1$ satisfies the fine-tuning condition for all values of $\nu\gtrsim 0.5$ so that fixing $a=3.1$ is just a conservative assumption.}.
\vspace{0.5cm}
In fact the allowed (fine-tuned) region which corresponds to $\delta<1$ ($\delta>1$) is above (below) the solid line. We see that in the limit $\nu\to\infty$ (where the RS metric is reached) the bound on $a$ corresponds to $a>2$ as in the case of AdS$_5$ metric.

The Higgs mass is given by Eq.~(\ref{muSM}) as
\begin{equation}
m_H^2=\frac{2}{Zc_K^2}\left(\frac{M_1}{k}-a  \right)m_{KK}^2
\label{HiggsmassSW}
\end{equation}
so that the required fine-tuning to reproduce the experimental value $m_H=126$ GeV depends on the values of $Z$ and $m_{KK}$. If we quantify the fine-tuning by imposing that $M_1=ak+\Delta M_1$ then
\begin{equation}
\frac{\Delta M_1}{M_1}=\frac{Zc_K^2}{2a}\, \frac{m_H^2}{m_{KK}^2}
\end{equation}
so that the larger (the smaller) $Z$ ($m_{KK}$) the less fine-tuned is the theory. In particular the factor $Z$ was one of the elements which increased the fine-tuning in the Higgs mass determination in the case of AdS$_5$ as $Z_{RS}<0.5$ for $a>2$. In the deformed metric of this section we can achieve values of $Z\gg 1$ which then lower the amount of fine-tuning. In particular in Fig.~\ref{fig3} we plot dashed contour lines of constant $Z= 0.75$ (on the right side), 1 (next curve, 2 (next curve) and 5 (on the left side). In the region allowed by the solution to the hierarchy problem (above the solid line) we see that, depending on the values of $a$ and $\nu$, we can have values of $Z$ as large as $\sim 5$. The second factor which determines the degree of fine-tuning in fixing a light Higgs is the value of $m_{KK}$ which is constrained by EWPT as we will see in the next section. In fact in the region where $Z\sim 5$ if $m_{KK}=\mathcal O(1)$ TeV, as we will see next, then $\Delta M_1/M_1=\mathcal O(1)$ and there is no fine-tuning at all.

To analyze perturbative unitarization of the theory one can compute the coupling of the Higgs to gauge bosons and in
particular one can prove that~\cite{CGQ2}
\begin{equation}
h_{WWH}^2=h_{WWH,SM}^2\left(1-\xi\right),\quad \xi=\mathcal O(m_H^2/m_{KK}^2)\simeq 0.01
\label{acoplo}
\end{equation}
so a light Higgs unitarizes the theory in a similar way as the SM Higgs.

\subsection{Electroweak constraints}
For the singular metric of Eq.~(\ref{A}) the electroweak observables of Eq.~(\ref{STYWfinal}) do not have an analytical expression. They have to be computed numerically. However in the region where $Z$ is large enough we can approximate $S$ and $T$ by
\vspace{0.5cm}
\begin{figure}[htb]
\centerline{\psfig{file=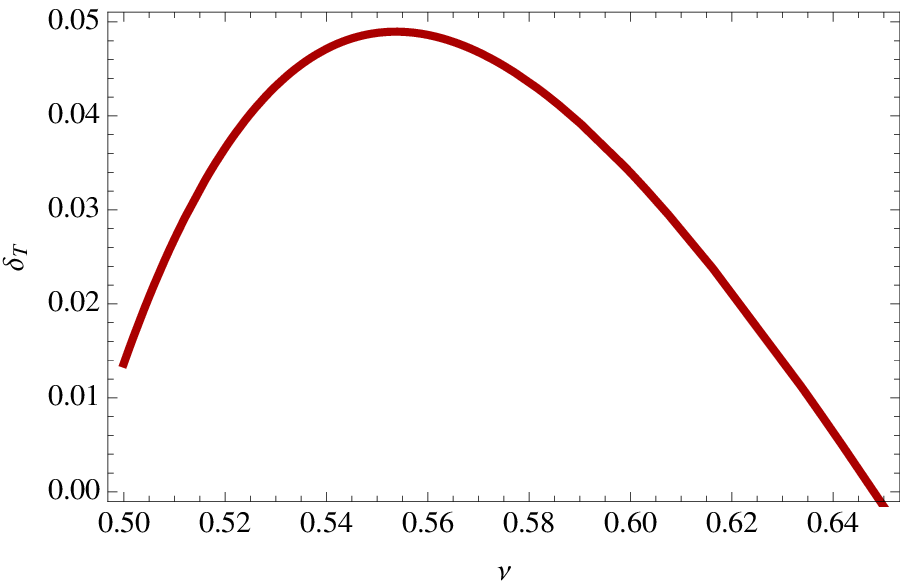,width=2.5in}
\psfig{file=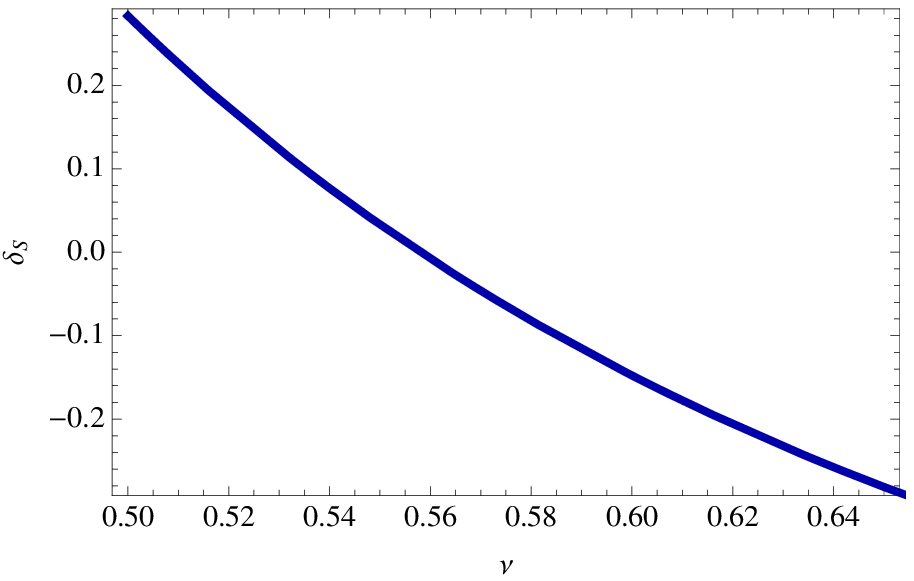,width=2.5in}}
\vspace*{8pt}
\caption{Left panel: $\delta_T$ as a function of $\nu$. Right panel: $\delta_S$ as a function of $\nu$. We have fixed $A(y_1)=35$ and $k\Delta=1$.
\protect\label{fig4}}
\end{figure}
\begin{equation}
\alpha T_{ap}\simeq s_W^2 ky_1 \frac{c_K^2}{Z^2}\,\frac{m_Z^2}{m_{KK}^2}\,,\quad
\alpha S_{ap}\simeq 8 s_W^2 c_W^2  \frac{c_K^2}{Z}\,\frac{m_Z^2}{m_{KK}^2}
\end{equation}
We will check the accuracy of the approximation by defining the corresponding relative errors as
\begin{equation}
\delta_T\equiv\frac{T_{ap}-T}{T},\quad \delta_S\equiv\frac{S_{ap}-S}{S} 
\end{equation}
which we plot in Fig.~\ref{fig4} as a function of $\nu$ for $k\Delta=1$ and $A(y_1)=35$. 
We can see from the left panel of Fig.~\ref{fig4} that $T_{ap}$ is a good approximation, within an error $\lesssim 5$ \%, while from the right panel we see that $S_{ap}$ approximates the exact value of $S$ only with errors $\lesssim 20$ \%. 
The reason for the different accuracy of both observables is that the goodness of the approximation is governed by the size of $Z$, and $T$ is suppressed by two powers of $Z$ while $S$ is suppressed only by one power of $Z$. The plots in Fig.~\ref{fig4} are valid for any value of $m_{KK}$.

\vspace{0.5cm}
\begin{figure}[htb]
\centerline{\psfig{file=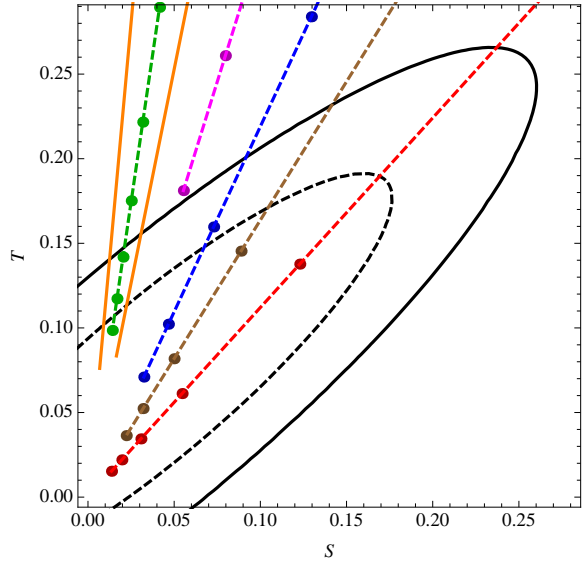,width=3.0in}}
\vspace*{8pt}
\caption{Predictions of the $S$ and $T$ parameters for the singular metric (\ref{A}) for $k\Delta=1$ and $a=3.1$. The dots on the dashed red line (on the right) correspond to $\nu=0.5$ and $m_{KK}\leq 3$ TeV with $\Delta m_{KK}=0.5$ TeV. The dots on the dashed grey line (next on its left) correspond to $\nu=0.525$ and $m_{KK}\leq 3$ TeV with $\Delta m_{KK}=0.5$ TeV. The dots on the dashed blue line (next on its left) correspond to $\nu=0.55$ and $m_{KK}\leq 3$ TeV with $\Delta m_{KK}=0.5$ TeV. The dots on the dashed magenta line (next on its left) correspond to $\nu=0.6$ and $m_{KK}\leq 3$ TeV with $\Delta m_{KK}=0.5$ TeV. Finally the dots on the dashed green line (on the left) correspond to $\nu=5$ and $m_{KK}\leq 12$ TeV with $\Delta m_{KK}=1$ TeV. The predictions of the RS model for $a=2$ and a localized Higgs are given by the solid orange lines. In all cases the larger value of $m_{KK}$ corresponds to the most inner dot.\protect\label{fig5}}
\end{figure}

In Fig.~\ref{fig5} we plot the values of the $S$ and $T$ parameters (using the exact numerical expressions) for $k\Delta=1$, $a=3.1$~\footnote{For which there is no hierarchy problem in the considered region of the parameter space, see Fig.~\ref{fig2}.} and various values of $\nu$, and we compare them with the region allowed by experimental data. Using the results from the plots in Fig.~\ref{fig5} we can see that the 95\% CL bounds on $m_{KK}$ considerably decrease with respect to their values in the RS model. In particular we can read from Fig.~\ref{fig5} the following bounds: $m_{KK}\gtrsim$ 0.8 TeV (for $\nu=0.5$), 1.2 TeV (for $\nu=0.525$), 1.8 TeV (for $\nu=0.6$) and 10 TeV (for $\nu=5$). In fact for the value $\nu=5$ we essentially recover the results from the RS model as we can see in the solid lines of Fig.~\ref{fig5} which correspond to the RS model for the cases of $a=2$ and a fully IR localized Higgs $a\to\infty$.

\begin{figure}[htb]
\centerline{\psfig{file=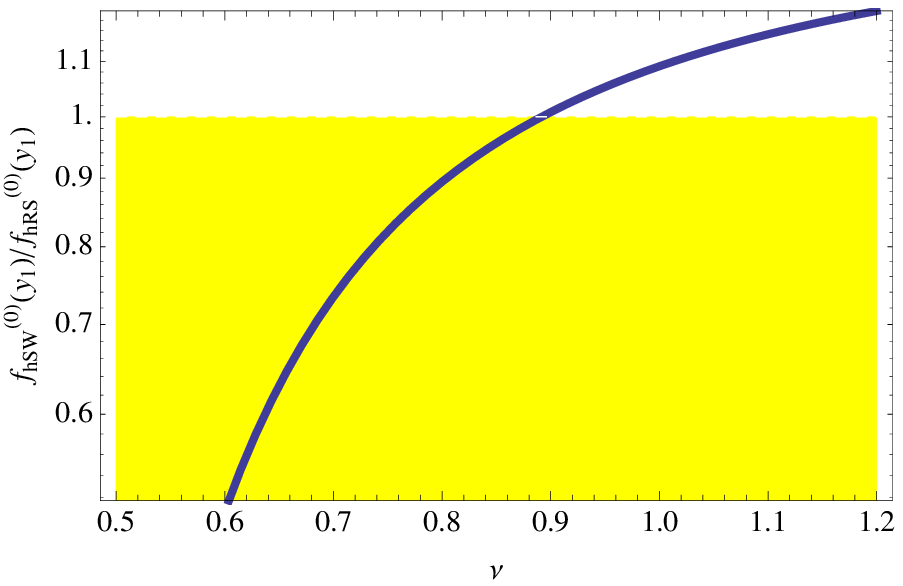,width=2.5in}
\psfig{file=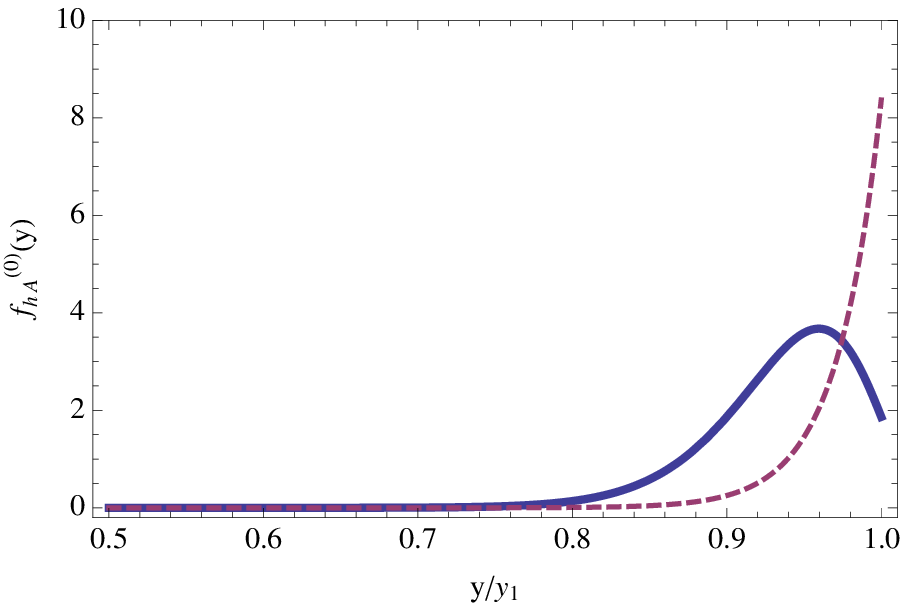,width=2.5in}}
\vspace*{8pt}
\caption{Left panel:  The ratio $f_{hSW}^{(0)}(y_1)/f_{hRS}^{(0)}(y_1)$ as a function of $\nu$, with $a=2$ in the AdS$_5$ metric and $a=3.1$, $k\Delta=1$ in the singular background metric. Right panel: Plot of $f_{hSW}^{(0)}$ (solid line) and $f_{hRS}^{(0)}$ (dashed line) as functions of $y/y_1$. We fix in both models $A(y_1)=35$. 
\protect\label{fig6}}
\end{figure}

We can understand the improvement from electroweak constraints in the soft-wall model by the different behaviour of the Higgs profile at the IR brane location $y_1$ (see e.g.~Ref.~\cite{Carmona:2011ib}). In fact the normalized physical Higgs wave function is defined as
\begin{equation}
f_{h\,A}^{(0)}(y) =N_0 e^{-A(y)}h(y),
\label{Higgsphys}
\end{equation}
where $N_0$ is a normalization factor such that
\begin{equation}
\int_0^{y_1}dy\left(f_{h\,A}^{(0)}(y)\right)^2=y_1
\label{normalizaci—n}
\end{equation}
As KK-modes are localized towards the IR brane it turns out that their overlapping with the Higgs excitations (and therefore their contribution to the electroweak observables $T$ and $S$) depends to a large extent on the values of the physical Higgs wave functions at the IR brane. In fact the values of the physical Higgs wave functions for the AdS$_5$ and singular background (SW) satisfy the inequality $f_{hSW}^{(0)}(y_1)<f_{hRS}^{(0)}(y_1)$ (for a fixed $y_1$) if the singular background metric departs sufficiently from AdS$_5$ at the IR brane. This property is exhibited in the left panel of Fig.~\ref{fig6} [the shadowed (yellow) region] where we plot $f_{hSW}^{(0)}(y_1)/f_{hRS}^{(0)}(y_1)$ as a function of $\nu$ for the fixed value $k\Delta=1$. We have used $a=2$ for the RS case while we are using $a=3.1$ for the SW case consistently with the previous comments about the solution of the hierarchy problem. In this way our result is a conservative one, as from Fig.~\ref{fig3} we can see that values $a<3.1$ would be consistent with the solution to the hierarchy problem for values $\nu>0.5$. Notice that the condition $f_{hSW}^{(0)}(y_1)<f_{hRS}^{(0)}(y_1)$ (shadowed region) is satisfied for $\nu<1$ where we expect deep deviations from AdS$_5$ near the IR brane. Notice also that in both cases we have fixed the condition $A(y_1)=35$ to solve the hierarchy problem so that $ky_1\leq 35$ for the SW case as we saw in Fig.~\ref{fig3}. The profiles of the physical Higgs wave functions $f_{hSW}^{(0)}(y)$ (for $\nu=0.5$, $k\Delta=1$, $A(y_1)=35$ and $a=3.1$) and $f_{hRS}^{(0)}(y)$ (with $ky_1=35$) are plotted as functions of $y/y_1$ in the right panel of Fig.~\ref{fig6} where we can see that $f_{hSW}^{(0)}(y)$ has a maximum at a value $y_{\rm max}<y_1$.

Implications at the LHC for this class of warped extra-dimensional models have been studied in Refs.~\cite{Carmona:2011ib,Antonio,Frank:2013qma}. These papers discuss both the production of electroweak and strong Kaluza-Klein gauge bosons. Only signals
involving the third generation of quarks seem to be feasible in order to probe this
scenario.

As we had tree-level contributions to the oblique parameters we did not pay too much attention for the moment to radiative corrections as they are expected to be subdominant, provided they are \textit{calculable}. This issue has been discussed in several papers~\cite{Delaunay:2010dw,Carmona:2011ib} and here we just review the main argument, which is purely based on dimensional analysis. We will exemplify this analysis with the $T$ and $S$ parameters. As there is no custodial symmetry protecting the $T$ parameter we can write the bulk effective Lagrangian terms as
\begin{equation}
\mathcal L_5=\frac{\kappa_T}{\Lambda_5^3}|H^\dagger D_M H|^2+
\frac{\kappa_S}{\Lambda_5^2}B^{MN}W_{MN}^a H^\dagger \sigma_a H
\end{equation}
where $\Lambda_5$ is the 5D cutoff scale and we assume canonical dimensions for the gauge fields, which can be done by defining the kinetic terms as $-(1/4g_5^2) \textrm{Tr} W_{MN}W^{MN}$ and $-(1/4g_5'^2)  B_{MN}B^{MN}$. The prefactors come just from the dimension of the Higgs propagating in the bulk $[H]=3/2$ and the dimensionless coefficients $\kappa_T$ and $\kappa_S$ are related to the $T$ and $S$ parameters. 

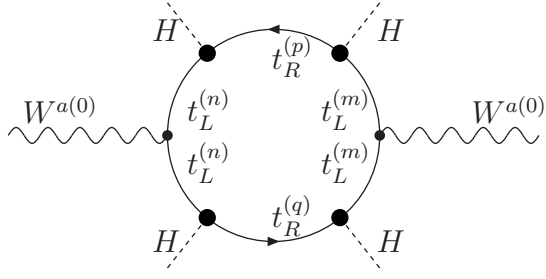
\begin{figure}[htb]
\begin{center} \begin{picture}(300,100)(0,0)
\ArrowArc(150,50)(40,0,180)
\ArrowArc(150,50)(40,180,360)
\Photon(50,50)(110,50){3}{5} \Vertex(110,50){2}
\Photon(190,50)(250,50){3}{5} \Vertex(190,50){2}
\DashLine(125,19)(110,0){2}
\DashLine(125,81)(110,100){2}
\DashLine(175,81)(190,100){2}
\DashLine(175,19)(190,0){2}
\Text(110,10)[]{$H$}
\Text(195,10)[]{$H$}
\Text(110,90)[]{$H$}
\Text(195,90)[]{$H$}
\GCirc(125,19){3}{0}
\GCirc(175,19){3}{0}
\GCirc(125,81){3}{0}
\GCirc(175,81){3}{0}
\Text(118,60)[l]{$t_L^{(n)}$}
\Text(118,40)[l]{$t_L^{(n)}$}
\Text(178,60)[]{$t_L^{(m)}$}
\Text(178,40)[]{$t_L^{(m)}$}
\Text(150,80)[l]{$t_R^{(p)}$}
\Text(150,20)[l]{$ t_R^{(q)}$}
\Text(70,60)[]{$W^{a(0)}$}
\Text(240,60)[]{$W^{a(0)}$}
\end{picture}  
\end{center}
\caption{One-loop contribution to the $T$-parameter.}
\label{fig7}
\end{figure}

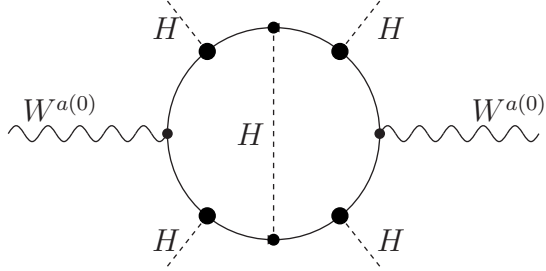
\begin{figure}[htb]
\begin{center} \begin{picture}(300,100)(0,0)
\ArrowArc(150,50)(40,0,180)
\ArrowArc(150,50)(40,180,360)
\Photon(50,50)(110,50){3}{5} \Vertex(110,50){2}
\Photon(190,50)(250,50){3}{5} \Vertex(190,50){2}
\DashLine(125,19)(110,0){2}
\DashLine(125,81)(110,100){2}
\DashLine(175,81)(190,100){2}
\DashLine(175,19)(190,0){2}
\DashLine(150,87)(150,13){2}
\GCirc(150,90){2}{0}
\GCirc(150,10){2}{0}
\Text(110,10)[]{$H$}
\Text(195,10)[]{$H$}
\Text(110,90)[]{$H$}
\Text(195,90)[]{$H$}
\Text(142,50)[]{$H$}
\GCirc(125,19){3}{0}
\GCirc(175,19){3}{0}
\GCirc(125,81){3}{0}
\GCirc(175,81){3}{0}
%\Text(118,60)[l]{$t_L^{(n)}$}
%\Text(118,40)[l]{$t_L^{(n)}$}
%\Text(178,60)[]{$t_L^{(m)}$}
%\Text(178,40)[]{$t_L^{(m)}$}
%\Text(150,80)[l]{$\tilde t_R^{(p)}$}
%\Text(150,20)[l]{$\tilde t_R^{(q)}$}
\Text(70,60)[]{$W^{a(0)}$}
\Text(240,60)[]{$W^{a(0)}$}
\end{picture}  
\end{center}
\caption{Two-loop contribution to the $T$-parameter.}
\label{fig8}
\end{figure}

The leading one-loop diagram contributing to the $T$ 
parameter is that given in Fig.~\ref{fig7} where the top quark is exchanged in the loop
%%%%%%%%
with a 5D coupling $y_t$ with dimension $[y_t]=-1/2$~\footnote{This comes from the bulk interaction term $y_t \int d^5 x \bar\psi_L H \psi_R+h.c.$ with $[\psi]=2$.}. The contribution from the diagram in Fig.~\ref{fig7} is then proportional to $y_t^4$ times the 5D integral over momenta. As $[y_t^4]=-2$ it follows that the term in the momentum integral proportional to $\eta_{\mu\nu}$ has dimension -1 and it is thus finite. If we subtract the top-quark zero mode the result should be controlled by the scale $\rho$~\cite{Carmona:2011ib}. This result only holds at one loop. In fact if we consider the two-loop diagram of Fig.~\ref{fig8}
we get its contribution to be proportional to $y_t^6$ and thus as $[y_t^6]=-3$ the momentum integral has dimension 0, and so it is logarithmically divergent (still calculable). Only the three loop diagram, proportional to $y_t^8$, will be linearly sensitive to the cutoff and thus non calculable but tiny.

A similar analysis can be done for the $S$ parameter. At one-loop it is generated by the diagram in Fig.~\ref{fig8} 
%%%%%%%%
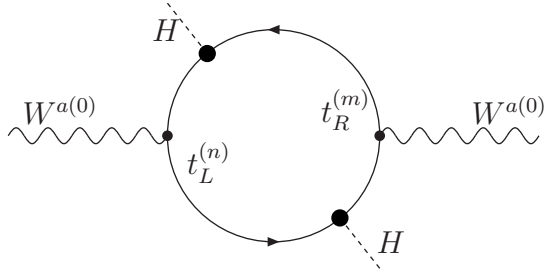
\begin{figure}[htb]
\begin{center} \begin{picture}(300,100)(0,0)
\ArrowArc(150,50)(40,0,180)
\ArrowArc(150,50)(40,180,360)
\Photon(50,50)(110,50){3}{5} \Vertex(110,50){2}
\Photon(190,50)(250,50){3}{5} \Vertex(190,50){2}
%\DashLine(125,19)(110,0){2}
\DashLine(125,81)(110,100){2}
%\DashLine(175,81)(190,100){2}
\DashLine(175,19)(190,0){2}
%\Text(110,10)[]{$H$}
\Text(195,10)[]{$H$}
\Text(110,90)[]{$H$}
%\Text(195,90)[]{$H$}
%\GCirc(125,19){3}{0}
\GCirc(175,19){3}{0}
\GCirc(125,81){3}{0}
%\GCirc(175,81){3}{0}
%\Text(118,60)[l]{$t_L^{(n)}$}
\Text(118,40)[l]{$t_L^{(n)}$}
\Text(178,60)[]{$t_R^{(m)}$}
%\Text(178,40)[]{$t_L^{(m)}$}
%\Text(150,80)[l]{$t_R^{(m)}$}
%\Text(150,20)[l]{$ t_R^{(q)}$}
\Text(70,60)[]{$W^{a(0)}$}
\Text(240,60)[]{$W^{a(0)}$}
\end{picture}  
\end{center}
\caption{One-loop contribution to the $S$-parameter.}
\label{fig9}
\end{figure}
which is proportional to $y_t^2$. As $[y_t^2]=-1$ it means that the momentum integral has dimension -1 and it is thus finite. At two-loop we consider the diagram of Fig.~\ref{fig10}
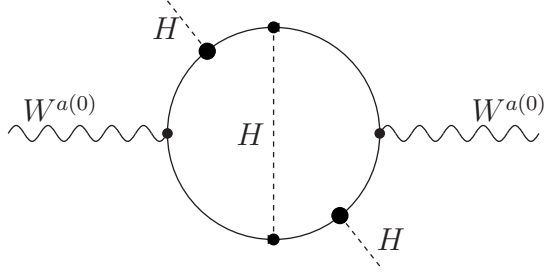
\begin{figure}[htb]
\begin{center} \begin{picture}(300,100)(0,0)
\ArrowArc(150,50)(40,0,180)
\ArrowArc(150,50)(40,180,360)
\Photon(50,50)(110,50){3}{5} \Vertex(110,50){2}
\Photon(190,50)(250,50){3}{5} \Vertex(190,50){2}
%\DashLine(125,19)(110,0){2}
\DashLine(125,81)(110,100){2}
%\DashLine(175,81)(190,100){2}
\DashLine(175,19)(190,0){2}
\DashLine(150,87)(150,13){2}
\GCirc(150,90){2}{0}
\GCirc(150,10){2}{0}
%\Text(110,10)[]{$H$}
\Text(195,10)[]{$H$}
\Text(110,90)[]{$H$}
%\Text(195,90)[]{$H$}
\Text(142,50)[]{$H$}
%\GCirc(125,19){3}{0}
\GCirc(175,19){3}{0}
\GCirc(125,81){3}{0}
%\GCirc(175,81){3}{0}
%\Text(118,60)[l]{$t_L^{(n)}$}
%\Text(118,40)[l]{$t_L^{(n)}$}
%\Text(178,60)[]{$t_L^{(m)}$}
%\Text(178,40)[]{$t_L^{(m)}$}
%\Text(150,80)[l]{$\tilde t_R^{(p)}$}
%\Text(150,20)[l]{$\tilde t_R^{(q)}$}
\Text(70,60)[]{$W^{a(0)}$}
\Text(240,60)[]{$W^{a(0)}$}
\end{picture}  
\end{center}
\caption{Two-loop contribution to the $S$-parameter.}
\label{fig10}
\end{figure}
which is proportional to $y_t^4$ and from $[y_t^4]=-2$ we see that the corresponding loop integral is logarithmically divergent and thus calculable. Again only the three loop diagrams, proportional to $y_t^6$, will be linearly sensitive to the cutoff but with a huge loop suppression.

Let us finally notice that these results are valid only for a bulk propagating Higgs. If the Higgs is localized on the IR brane, and so it has mass dimension $[H]=1$, then we can write the tree-level effective Lagrangian as
\begin{equation}
\mathcal L_4=\frac{\kappa_T}{\Lambda_4^2}|H^\dagger D_\mu H|^2+
\frac{\kappa_S}{\Lambda_4^2}B^{\mu\nu}W^a_{\mu\nu}H^\dagger \sigma_aH
\end{equation}
and the localized top Yukawa coupling $h_t$~\footnote{From the brane localized Yukawa interaction $h_t\int d^4x \bar\psi_L H_{loc}\psi_R+h.c.$ with $[\psi_{L,R}]=2$.} has dimension $[h_t]=-1$. In this case the diagram of Fig.~\ref{fig7} contributing to the $T$ parameter is proportional to $h_t^4$ with dimension $[h_t^4]=-4$ such that the term proportional to $\eta_{\mu\nu}$ in the calculation of the diagram has dimension +2, which means a quadratic sensitivity to the cutoff $\Lambda_4$ and the corresponding contribution not being computable. Likewise the diagram of Fig.~\ref{fig9} contributing to the $S$ parameter is proportional to $h_t^2$ with dimension $[h_t^2]=-2$, which means a logarithmic sensitivity to the scale of the $S$ parameter, while the two loop diagram of Fig.~\ref{fig10} proportional to $h_t^4$, with dimension $[h_t^4]=-4$, yields a quadratic sensitivity.

%%%%%%%%%%

In this section we have seen how in models with the Standard Model propagating in the bulk of the extra dimension we can cope with the experimental values of the electroweak observables by departing from the AdS$_5$ geometry near the IR brane. Of course in these models we can also fix the quark and lepton structure of masses and mixing angles by the different localization of fermions in the bulk by means of different 5D Dirac masses. As in the RS model this process creates FCNC and CP violating operators which should be made consistent with their experimental bounds~\cite{CGQ5,Carmona:2011ib,vonGersdorff:2012tt}. This issue will not be considered here, as it will be reviewed elsewhere~\cite{Gero}.

\section{Custodial Models}
Another way of enforcing the strong constraints imposed by the $T$ parameter is to promote the custodial symmetry to a gauge symmetry in the bulk. This idea was put forward in Ref.~\cite{Agashe:2003zs}.
In particular the construction proposed there consisted in enlarging the electroweak gauge group in the bulk to $SU(2)_L\otimes SU(2)_R\otimes U(1)_\mathcal X$ (where $\mathcal X\equiv B-L$). The gauge group is then broken by boundary conditions in the UV brane to the Standard Model gauge group $SU(2)_L\otimes U(1)_Y$ (where $Y=\mathcal X+T_R^3$) while it remains unbroken in the IR brane. We can thus impose Neumann ($N$) boundary conditions on the $y_\alpha$ brane [$f'_A(y_\alpha)=0$] to keep the corresponding gauge boson massless and Dirichlet ($D$) boundary conditions [$f_A(y_\alpha)=0$] to give it an (infinite) mass. Thus the required boundary conditions on the (UV, IR) branes at $y=(0,y_1)$ are~\footnote{All gauge couplings here are 5D. Their 4D counterparts are defined according to the usual formula: $g_5=g_4\sqrt{y_1}$.}
\begin{equation}
\begin{array}{rrc}
&W_L^{1,2,3} & =(N,N) \\
&W_R^{1,2}  &  =(D,N)\\
\mathcal Z&=\frac{1}{\sqrt{g_R^2+g_\mathcal X^2}}\left[g_R W_R^3 - g_\mathcal X \mathcal X \right]& =(D,N)\\
  B&=\frac{1}{\sqrt{g_R^2+g_\mathcal X^2}}\left[g_R W_R^3 + g_\mathcal X \mathcal X \right]& =(N,N)
\end{array}
\end{equation}
and the 5D covariant derivative becomes in terms of the new gauge fields
\begin{equation}
D_M=\partial_M-i\left(g \sum_{a=1,2,3}W_M^a T^a_L+g_R \sum_{i=1,2}W^{i}_{R\,M} T_R^i+g_{\mathcal Z}\mathcal Z_M Q_{\mathcal Z}+g'B_M Y\right)
\end{equation}
where $g$ and $g'=g_Rg_\mathcal X/\sqrt{g_R^2+g_\mathcal X^2}$ are the $SU(2)_L\otimes U(1)_Y$ couplings, $g_R$ is the $SU(2)_R$ coupling, $g_{\mathcal Z}=\sqrt{g_R^2+g_\mathcal X^2}$ the $\mathcal Z$ coupling  and $Q_{\mathcal Z}=T_R^3-\sin^2\theta \,Y$ is the $\mathcal Z$ charge with the $W_R^3-\mathcal X$ mixing angle $\theta$ given by
\begin{equation}
\sin\theta=\frac{g_\mathcal X}{g_{\mathcal Z}}
\end{equation} 

Since we have enlarged the bulk gauge symmetry the right-handed fermions should be promoted to doublets under $SU(2)_R$. However since we are breaking this symmetry through the UV boundary conditions only one of the components of the doublet should have a zero mode while the other should not have any. This means that to have two zero modes we need to double the number of doublets with different boundary conditions. For instance for the right-handed up quarks we will need to $SU(2)_R$ doublets
\begin{equation}
Q_{R\,1}=\left(\begin{array}{c} u_R\\ \tilde d_R\end{array}\right),\quad
Q_{R\,2}=\left(\begin{array}{c} \tilde u_R\\  d_R\end{array}\right)
\end{equation}
where only the untilded fields possess a zero mode. In fact since we are breaking $SU(2)_R$ through the orbifold boundary conditions, one component of $SU(2)_R$ must be even (with a zero mode) and the other odd (no zero mode). This doubling on the number of fields is only required in the quark sector as in the lepton sector for the $SU(2)_R$ doublet
\begin{equation}
L_R=\left(\begin{array}{c} \ell_R\\ \tilde \nu_R\end{array}\right)
\end{equation}
only the $\ell_R$ field needs to possess a zero mode.

Finally the Higgs is considered as a bidoublet under $SU(2)_L\otimes SU(2)_R$, with a potential $V(H)$ localized at the IR brane such that it provides a minimum at $v=246$ GeV and with a small ratio to the warped down curvature scale $v/\rho$. The Yukawa couplings to fermions are written by the IR localized terms $H(y_{5\,t}Q_LQ_{R\,1}+y_{5\,b}Q_LQ_{R\,2}+y_{5\,\tau} L_LL_R)$.

As the KK-modes are localized towards the IR boundary and both the bulk and the IR boundary respect exactly the custodial symmetry (even for $g'\neq 0$) the vanishing of the $T$ parameter is enforced by the custodial symmetry. In fact the detailed calculation~\cite{Agashe:2003zs} shows that there is a \textit{tree-level} cancellation between the contribution from $W_L$ and $W_R$ and the terms enhanced by $ky_1$ vanish. This yields at the tree level, and assuming that light fermions are localized at the UV brane, the following expressions for $S$ and $T$ 
\begin{eqnarray}
S&\simeq& 2\pi c_K^2 \frac{v^2}{m_{KK}^2}+\mathcal O\left(\frac{v^4}{m_{KK}^4} \right)\nonumber\\
T&\simeq & 0\ .
\label{ese}
\end{eqnarray}
where $c_K\simeq 2.45$.

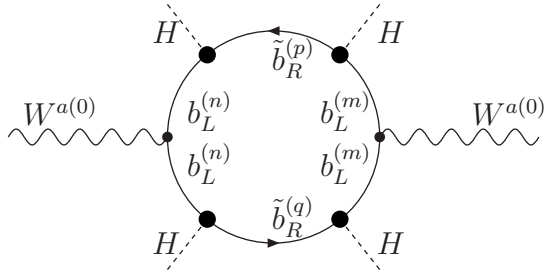
\begin{figure}[htb]
\begin{center} \begin{picture}(300,100)(0,0)
\ArrowArc(150,50)(40,0,180)
\ArrowArc(150,50)(40,180,360)
\Photon(50,50)(110,50){3}{5} \Vertex(110,50){2}
\Photon(190,50)(250,50){3}{5} \Vertex(190,50){2}
\DashLine(125,19)(110,0){2}
\DashLine(125,81)(110,100){2}
\DashLine(175,81)(190,100){2}
\DashLine(175,19)(190,0){2}
\Text(110,10)[]{$H$}
\Text(195,10)[]{$H$}
\Text(110,90)[]{$H$}
\Text(195,90)[]{$H$}
\GCirc(125,19){3}{0}
\GCirc(175,19){3}{0}
\GCirc(125,81){3}{0}
\GCirc(175,81){3}{0}
\Text(118,60)[l]{$b_L^{(n)}$}
\Text(118,40)[l]{$b_L^{(n)}$}
\Text(178,60)[]{$b_L^{(m)}$}
\Text(178,40)[]{$b_L^{(m)}$}
\Text(150,80)[l]{$\tilde b_R^{(p)}$}
\Text(150,20)[l]{$\tilde b_R^{(q)}$}
\Text(70,60)[]{$W^{a(0)}$}
\Text(240,60)[]{$W^{a(0)}$}
\end{picture}  \end{center}
\caption{One loop diagram contributing to the $T$-parameter.}
\label{diagram}
\end{figure}

\begin{figure}[htb]
\centerline{\psfig{file=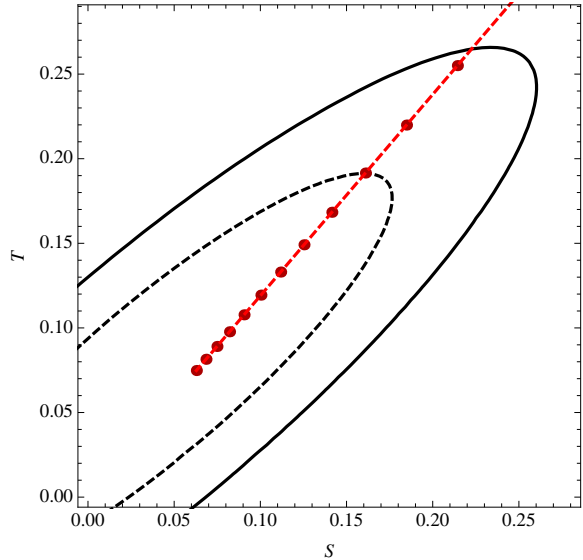,width=3.0in}}
\vspace*{8pt}
\caption{Prediction of the $S$ and $T$ parameters in the custodial model of this section for values of $m_{KK}\leq 6 $ TeV, step $\Delta m_{KK}=0.25$ TeV. The most inner dot corresponds to $m_{KK}=6$ TeV.
\protect\label{fig12}}
\end{figure}

At the one-loop level, on top of the usual Standard Model leading contribution from the top quark zero mode,   
the dominant contribution comes from a loop where the KK-modes $(t_L^{(n)},t_R^{(n)})$ and $(b_L^{(m)},\tilde b_R^{(m)})$ are exchanged, as in the diagrams of Figs.~\ref{fig7} and \ref{diagram} where big dots correspond to Higgs insertions   $\langle H\rangle =v/\sqrt{2}$.
In fact after summation over all KK-modes in the loop, and considering $c_{t_L}\sim 0.4$, Ref.~\cite{Agashe:2003zs} gets the numerical result
\begin{equation}
T\simeq 90 \frac{m_t^2}{m_{KK}^2}\ .
\label{te}
\end{equation}
where $m_{KK}$ is the mass of the first KK-mode for gauge bosons. In Fig.~\ref{fig12} we plot the values of the $S$ and $T$ parameters
%
%\vspace{0.5cm}
from Eqs.~(\ref{ese}) and (\ref{te}) and we compare them with the experimental data on electroweak observables. The dots correspond to values of $m_{KK}\leq 6$ TeV with an interval between points of $\Delta m_{KK}=0.25$ TeV. From there one can infer the lower bound of $m_{KK}\gtrsim 3$ TeV at the 95\% CL.

Up to this point we considered the Higgs as a fundamental field but there are theories where the Higgs is not introduced as an extra matter because:
\begin{itemize}
\item
Either, it is a condensate of a pair of fermion zero modes induced by the KK gluons~\cite{Dobrescu:1998dg,Burdman:2007sx,Bai:2008gm}. 
\item
Or the Higgs is the fifth component of the gauge bosons of an enlarged bulk gauge symmetry.
\end{itemize} 
In these cases EWSB is generated dynamically. In particular the latter possibility is being explored in depth at present and it is dubbed Gauge-Higgs Unification (GHU)~\cite{H1,H2,H3,H4,H5}.

GHU is actually an alternative to supersymmetry~\cite{Antoniadis:2001cv}, where the (5D) gauge symmetry in the bulk $\mathcal G$ protects the mass of extra-dimensional components of gauge bosons~\cite{vonGersdorff:2002as,vonGersdorff:2002rg,vonGersdorff:2002us}. This solution to the hierarchy problem requires an extended gauge group with respect to the SM gauge group. It can be constructed in flat or warped space, although in warped space the GIM-RS mechanism protects the theory with differently localized fermion fields from huge flavor violation, which otherwise would require severe constraints on the mass of KK modes~\cite{Delgado:1999sv}. The four-dimensional components of 5D gauge bosons $(A_\mu^a)$ of $\mathcal G$ with $(N,N)$ boundary conditions span the four-dimensional gauge group while the fifth components $(A_5^{\hat a})$, with $(N,N)$ boundary conditions~\footnote{Correspondingly $A_5^a$ and $A_\mu^{\hat a}$ have $(D,D)$ boundary conditions.}, contain the four-dimensional Higgs (and Goldstone boson) fields in a number equal to the number of Pseudo Goldstone Bosons (PGB) which are left out in the four dimensional theory. 

In general $\mathcal G$ will be broken by boundary conditions to $\mathcal H_{UV}$ ($\mathcal H_{IR}$) on the UV (IR) brane. For $\mathcal H_{UV}=SU(2)_L\otimes U(1)_Y$ the number of PGB is $\dim(\mathcal G/\mathcal H_{IR})$ so different models differ by different choices for $\mathcal G$ and $\mathcal H_{IR}$. Some models~\cite{GHU1,GHU2,GHU3} are defined in the table below.
\vspace{.5cm} 
% \begin{table}
  \begin{center}
\begin{tabular}{||c|c||}
\hline\noalign{\smallskip}
Model & \# PGB ($A_5^{\hat a}$)  \\
\hline
{SO(4)}/{SO(3)} & 6-3=3 (Higgsless SM)\\
{SU(3)}/{SU(2)$\otimes$U(1)} & 8-4=4$(H_{SM})$\\ 
{SO(5)}/{SO(4)} & 10-6=4 $(H_{SM})$\\
{SO(6)}/{SO(5)} & 15-10=5 ($H_{SM}$ + singlet)\\
{SO(6)}/{SO(4)$\otimes$SO(2)} & 15-6-1=8 $(H_u,H_d)$\\
\noalign{\smallskip}\hline
\end{tabular}
\end{center}
%\label{tabla}
%\end{table}
%
\vspace{.5cm} The first model in the table, $SO(4)$, contains the SM Goldstone bosons and thus it is a Higgless model. This kind of models have been excluded by the discovery at the CERN LHC of the Higgs-like boson with mass $m_H\sim 126$ GeV. The second model presented in the table,
$SU(3)/SU(2)\otimes U(1)$, is sometimes studied as a toy model as it contains the SM Higgs on top of the SM Goldstone Bosons. In order to accommodate quark hypercharge it has to be accompanied by $U(1)$ factor(s). It does not contain the custodial symmetry on the IR brane and thus the $T$ parameter is not protected for it. The rest of the models in the table contain the custodial $SO(4)$ subgroup on the IR brane and so their contribution to the $T$ parameter is protected. In particular the third model on the table, $SO(5)$,  [usually accompanied by a common factor $U(1)_{B-L}$] is the minimal model which contains the custodial subgroup and thus it is commonly dubbed as Minimal Composite Higgs Model (MCHM)~\cite{GHU2}. 

In the dual theory $\mathcal G/\mathcal H_{IR}$ is characterized by the spontaneous breaking scale $f_h$ such that the expansion parameter in the theory is $\xi$
\begin{equation}
\xi\equiv\left(\frac{v}{f_h}\right)^2\quad \left\{ \begin{array}{l} \xi\to 0 \Rightarrow \textrm{SM limit} \\  \xi\to 1 \Rightarrow \textrm{Technicolor limit}\end{array}\right.
\end{equation}
The $\xi$ parameter controls perturbative unitarity through a relation similar to Eq.~(\ref{acoplo}). However unlike in the models presented in previous sections with a fundamental scalar Higgs, where the parameter $\xi\ll 1$, in the models presented in this section $\xi$ depends on $f_h$ and can thus be considered as a free parameter. For instance in the limit $\xi\to 0$ the SM result is obtained and the Higgs unitarizes the theory without the need of any extra particle. On the other extreme in the Technicolor limit $\xi\to 1$ all unitarity must be provided by new TeV resonances at scales close to the electroweak scale. For intermediate values of $0<\xi<1$ unitarity must be partially restored by resonances at scales which depend on the value of $\xi$. 

\begin{figure}[htb]
\centerline{\psfig{file=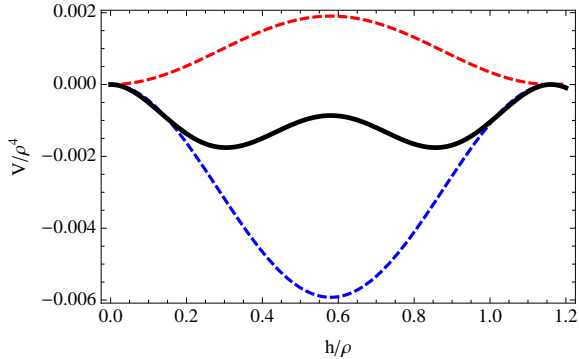,width=3.0in}}
\vspace*{8pt}
\caption{Potential (\ref{CWanal}) for a toy model. Upper dashed (red) line is the contribution from a gauge boson. Lower dashed (blue) line is the contribution from a fermion. Solid line is the total contribution from gauge bosons and fermions. This plot does \textit{not} correspond to any realistic model.\protect\label{fig13}}
\end{figure}

As we are identifying the Higgs field with the fifth component of a gauge field, 5D gauge invariance prevents any bulk potential term at the tree level. However a Coleman-Weinberg potential is generated at one-loop level which can be responsible for EWSB. This potential can be written on general grounds in the usual way
\begin{equation}
V=\frac{1}{2}\sum_n\int \frac{d^4p}{(2\pi)^4}\log\left[p^2+m_n^2(h)\right]
\label{CW}
\end{equation}
where $m_n$ are the mass eigenvalues of KK modes of the different bulk fields in the presence of the Higgs background $h$. The potential (\ref{CW}) was computed and given an analytic approximation in Ref.~\cite{Falkowski:2006vi} using holographic methods in arbitrary gravitational backgrounds. In particular, in the AdS$_5$ background it can be written as
\begin{equation}
V\simeq \sum_r \frac{N_r}{16\pi^2}\int dp \,p^3 \log\left[1+\frac{g_r^2f_h^2}{\rho^2}\frac{\sin^2(\lambda_r h/f_h)}{\sinh^2(p/\rho)}\right]
\label{CWanal}
\end{equation}
where $N_r$ are the degrees of freedom for the different fields ($N_r=3$ for gauge bosons, $N_r$=-4 for fermions) and $g_r$ the corresponding couplings ($g_r=g$ for gauge bosons, $g_r=y_t$ for the top quark). As for $\lambda_r$ they are representation dependent order one constants: $(\lambda_r h)^2$ are eigenvalues of the symmetric mass matrix $\mathcal M^{ab}=A_5^{\hat a} A_5^{\hat b} f^{a\hat a\hat c}f^{b\hat b \hat c}$ and the Higgs decay constant $f_h$ is given by
\begin{equation}
g f_h=\sqrt{\frac{2}{ky_1}}\rho
\end{equation}
We can see that the potential (\ref{CWanal}) is periodic. The gauge components have a minimum at $h=0$ while the fermionic components have a maximum at $h=0$ and thus trigger EWSB and a minimum at $\sin(\lambda_r h/f_h)=1$, i.e.~$\lambda_r h/f_h=\pi/2$. Depending on the number of bosonic and fermionic degrees of freedom one can generate a non-trivial minimum in between. This situation has been exemplified in the plot of Fig.~\ref{fig13} where we have considered a case with $\lambda_r=1$ ($\forall r$) and the number of bosonic and fermionic degrees of freedom as required to generate such an intermediate minimum. 

Moreover one can consider in general an effective theory~\cite{Espinosa:2010vn} parametrized by $\xi$, which measures the degree of compositeness of the Higgs. In this theory all Higgs couplings (cubic, quartic, $HWW$, \dots) depart from the SM values by quantities which are proportional to $\xi$. These models have been confronted to EWPT and direct searches~\cite{Espinosa:2012qj}. The former ones provide the strongest constraints which yield typical bounds $\xi< 0.18$ at 99\% CL in the absence of additional contributions to the $S$ and $T$ parameters. 

\section{Conclusions}
It is clear at this moment that the next step in Higgs physics belongs to the LHC, in particular to the ATLAS and CMS collaborations, so that confronting in great accuracy different theories on electroweak symmetry breaking with experimental data should wait till LHC13-14, when more statistics will be accumulated and Higgs production rates will be given more accurately. In particular if theories with extra dimensions are \textit{the} solution to the hierarchy problem the Higgs is expected to be composite to some extent, which should leave imprints in experimental data and in particular on Higgs production strengths. In any case if a warped space is a solution of the hierarchy problem it requires a UV completion as the theory becomes strongly coupled at some multi-TeV scale.
But even if it turns out that the detection of the degree of Higgs compositeness be outside the LHC accuracy it is clear that a confirmation of the extra dimensional nature of the space should be made by direct detection of the extra states that all these theories predict: KK excitations, extra 4D matter (as the extra states required by the custodial models), \dots. So the main (obvious) conclusion is that from 2015 on, when LHC13-14 will restart operations, we should stay tuned.

\section*{Acknowledgments}

Work partly supported by the Spanish Consolider-Ingenio 2010 Programme CPAN (CSD2007-00042) and by CICYT-FEDER-FPA2011-25948.

\end{document}